\documentclass[aps,prl,twocolumn,showpacs,preprintnumbers,amsmath,amssymb,superscriptaddress]{revtex4}
\usepackage{graphicx}
\usepackage{bm}

\begin{document} 

\title{Bona Fide Thermodynamic Temperature in Nonequilibrium Kinetic
Ising Models}

\author{Francisco Sastre}
\affiliation{Instituto de F\'{\i}sica de la Universidad de Guanajuato, AP E-143,
CP 37150, Le\'on, Gto., M\'exico}
\affiliation{CEA -- Service de Physique de l'\'Etat Condens\'e,~CEN~Saclay,~91191~Gif-sur-Yvette,~France}

\author{Ivan Dornic}
\affiliation{CEA -- Service de Physique de l'\'Etat Condens\'e,~CEN~Saclay,~91191~Gif-sur-Yvette,~France}

\author{Hugues Chat\'e}
\affiliation{CEA -- Service de Physique de l'\'Etat Condens\'e,~CEN~Saclay,~91191~Gif-sur-Yvette,~France}

\date{\today}

\begin{abstract}
We show that a nominal temperature
can be consistently and uniquely defined everywhere in the
phase diagram of large classes of
nonequilibrium kinetic Ising spin models.
In addition, we confirm the recent proposal
that, at critical points, the large-time ``fluctuation-dissipation ratio''
$X_\infty$ is a universal amplitude
ratio and find in particular
$X_\infty\approx 0.33(2)$ and $X_\infty=\frac{1}{2}$
for the magnetization in, respectively, the two-dimensional Ising
and voter universality classes. 
\end{abstract}

\maketitle

The fluctuation-dissipation theorem (FDT), which relates 
the correlation and  response (or susceptibility) functions
during the return to equilibrium following a small perturbation,
may be used to provide an absolute definition of the temperature 
of a physical system.
However, there are many instances, 
such as glassy materials and coarsening systems,
 where  the standard form of the FDT breaks down, 
because the equilibration timescales are astronomical or even infinite
in the thermodynamic limit.
Relaxation properties then depend upon both $t$, the time
at which perturbations
 are applied, and $t'$, at which measurements are taken, 
giving rise to ``aging phenomena''.
In this context, guided by 
the dynamics of some
mean-field spin glass models, Cugliandolo and Kurchan \cite{CUKU}
have proposed a generalized form of the FDT:
\begin{equation}
\label{def_X}
R(t,t')=\frac{X(t,t')}{T} \frac{\partial C(t,t')}{\partial t},
\end{equation}
where the two-time response and  auto-correlation functions  $R(t,t')$
and $C(t,t')$ of some physical observable,
 as well as the so-called FDT ratio $X(t,t')$,
are not functions of the $t'-t$,
and, moreover, $X$ depends functionally on $C$.
The limit value $X_{\infty}$ often taken by $X$ at large times 
is usually interpreted as the emergence of an  ``effective
temperature'' $T_{\rm eff}= T/X_\infty$ \cite{CUKUPE_PRE97}.

The picture suggested by the above generalized FDT scenario
has received a large number of confirmations \cite{REVIEW_LETICIA},
both from experimental \cite{EXPERIMENTS} and 
numerical \cite{STRUCTURAL,GRANULAR,SHEAREDFLUIDS} studies 
of structural and spin glasses, granular matter, and gently sheared fluids,
to quote a few examples.
In these situations, the physical systems under consideration, 
though they may never achieve equilibration, 
evolve subject to a temperature $T$.
However, this ``nominal'' temperature is generally not even defined
for {\it intrinsically} nonequilibrium systems, 
e.g. stochastic models for which detailed balance is violated, 
although such systems are known to behave much
like their equilibrium counterparts, including critical behavior
and non-stationary properties of their two-time
autocorrelation and response functions.

In this Letter, we show that a nominal temperature
can be consistently and uniquely defined everywhere in the
phase diagram of large classes of
nonequilibrium kinetic Ising models.
We explain how both a ``dynamical'' definition of temperature
(using Eq.~(\ref{def_X})) and a ``geometrical'' approach
(in terms of the density of states) lead to an ambiguity
that can be lifted using a maximum-entropy argument.
This results in a unique,
{\it bona fide}  thermodynamic temperature
which takes the same value for different observables
and coincides with the usual one
when detailed balance is satisfied.
In addition, considering the critical points of our models,
we confirm the proposal \cite{GODRECHE_1,GODRECHE_2}
that $X_\infty$ is, in this context, a universal amplitude
ratio  of  dynamic scaling functions and find in particular
$X_\infty[M]\approx 0.33(2)$ and $X_\infty[M]=\frac{1}{2}$
for the magnetization $M$ in, respectively, 
the two-dimensional Ising 
and voter \cite{VOTPRL} classes. 

\begin{figure}
\includegraphics[width=7.5cm]{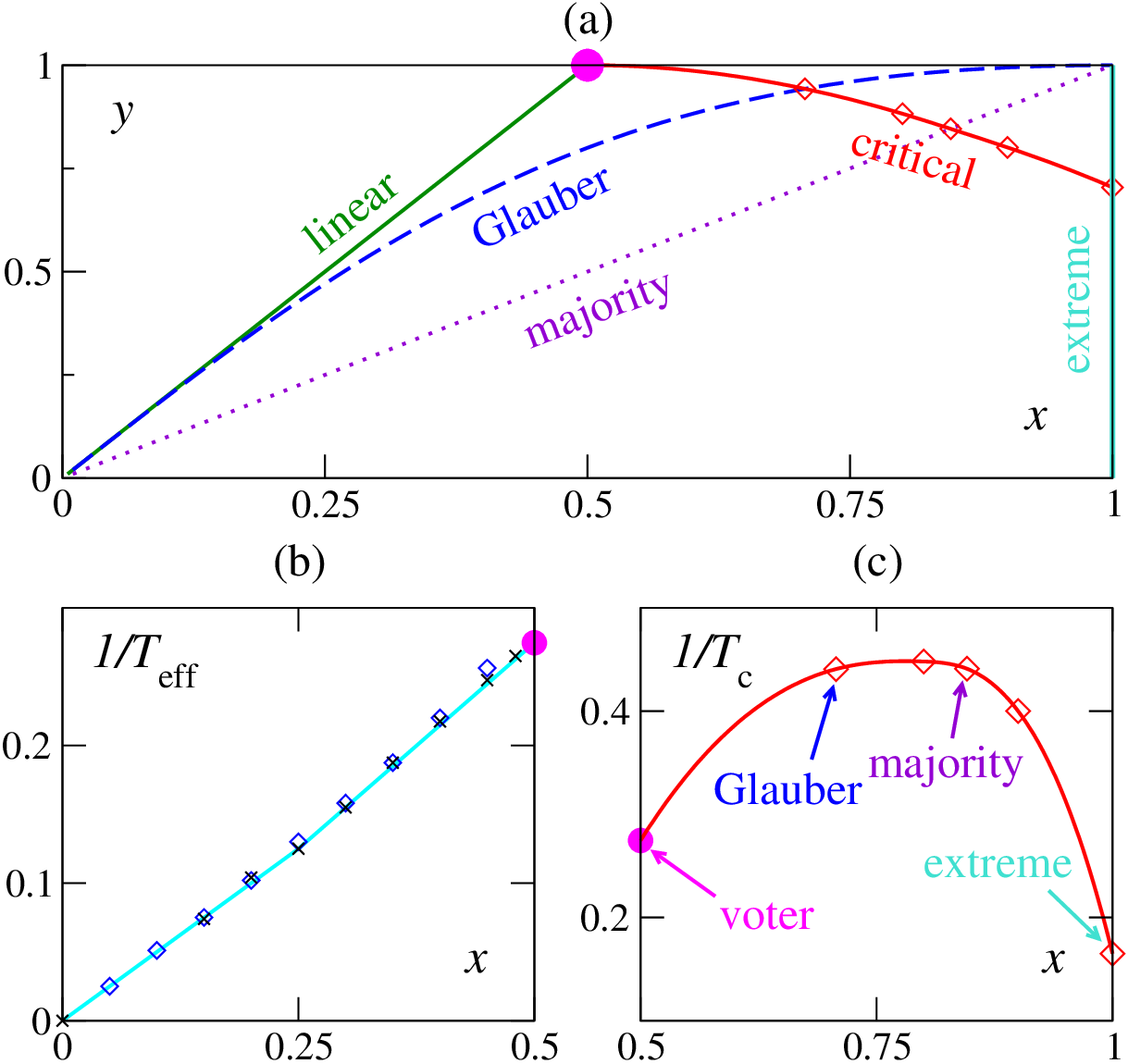}
\caption{(a): phase diagram in the $(x,y)$-plane: the Ising-critical line
separating the disordered region (below) from the ordered one (above)
terminates at the voter critical point (big filled circle).
(b): $\beta_{\rm dyn}$ (crosses) and $\beta_{\rm geo}$ (diamonds)
along the linear model line.
(c) Inverse effective temperature (computed from $\beta_{\rm dyn}$,
exact for Glauber and voter) along
the critical line.}
\label{f1}
\end{figure}

For simplicity, we illustrate our approach  
by considering the family of two-dimensional nonequilibrium kinetic 
spin models introduced some time ago in \cite{OLIVEIRA}
which are defined by the following
evolution rules. During an elementary
timestep, an Ising-like spin ($\sigma_r=\pm 1$) on a square lattice
is randomly picked up, and flipped with a probability 
$W[E_r=-\frac{1}{2}\sigma_r H_r]$ where $E_r\in\{-2,-1,0,1,2\}$ 
is the local (pseudo-)energy,
$H_r=\sum_{\mu=1}^{4}\sigma_{r+e_\mu}$ being the local (Weiss-like) field 
calculated over the four nearest neighbors of $\sigma_r$.
The local $Z_2$-symmetry of the dynamical rules is enforced
by demanding that  $W(-E)=1-W(E)$ (the system being homogeneous,
we drop spatial indexes whenever possible) 
and thus $W(0)=\frac{1}{2}$, leaving a two-parameter family
defined by $W(1) = \frac{1}{2}(1+x)$ and
$W(2) = \frac{1}{2}(1+y)$, with ---to favor ferromagnetic  
ordering--- $0 \le x,y \le 1$.
An alternative convenient parameterization \cite{DROUFFE} is to match 
the flip rate $W$  with the corresponding expression for 
Glauber dynamics ($W=\frac{1}{2}(1+\tanh{2\beta E})$ 
at temperature $T=1/\beta$).
This is only possible if one introduces  
two parameters 
$\beta_1 \equiv \beta_{E=1}$ and
$\beta_2 \equiv \beta_{E=2}$,  
such that $x=\tanh{2 \beta_1}$ and $y=\tanh{4 \beta_2}$, 
which  measure respectively  the strength of interfacial
 and bulk noise \cite{NOTE00}.
The  usual Glauber dynamics then corresponds
to $\beta_1=\beta_2=1/T$, or, equivalently, $y=2 x/(1+x^2)$.
Other well-studied models in the $(x,y)$-plane comprise
the majority model ($y=x$), the ``noisy voter'' or linear model
($y=2 x$), and the ``extreme'' model ($x=1$). 
Except on the Glauber line,
models in the $(x,y)$-plane do not obey detailed balance 
or possess an underlying  short-range Hamiltonian. 
Nevertheless, in agreement with a long-lasting
 conjecture backed up by field-theoretic 
arguments
\cite{GRINSTEIN85,VIRGINIATECH},
there exists a critical line $y_{\rm c}(x)$ 
separating a disordered (paramagnetic) from an ordered 
(ferromagnetic) phase and along which Ising static and 
dynamic critical exponents are numerically 
found (Fig.~\ref{f1}a) \cite{OLIVEIRA,EXPONENTSP1P2,DROUFFE}.
This line terminates at the voter model critical point $(\frac{1}{2},1)$,
across which the transition occurs in the absence of bulk
noise ($y=1$) \cite{VOTPRL}.

The gist of our argument to define
an effective temperature is the observation that, even 
if the FDT may  be broken at  large times, 
there always exists, for small time-differences ($t'-t \le {\cal{O}}(1)$
and $t \gg 1$), an equilibrium-like regime
 where the standard form (\ref{def_X}) is valid with $X= 1$. 
Thus, in this two-time sector, 
and in particular for $t'-t \ll 1$,
we view (\ref{def_X}) as a means of obtaining a dynamical
{\it definition} of an effective temperature \cite{NOTEXX}:
\begin{equation}
\label{def_Tdyn}
\frac{1}{T_{\rm dyn}}\equiv
\lim_{t \to \infty,t'-t \to 0}\frac{R(t,t')}{\partial_t C(t,t')} \;.
\end{equation}
To really make sense of  (\ref{def_Tdyn}), 
we must of course specify the response function $R(t,t')$,
i.e. we have to implement, for a generic $(x,y)$-model, 
the effect of a symmetry-breaking external field $h_r$ linearly conjugated 
to the local spin variable $\sigma_r$.
Probably the simplest choice
is to make the substitution 
$\sigma_r H_r \to \sigma_r (H_r+h_r)$ 
in the original expression of $W_r$, this simple recipe being also
consistent with the natural redefinition
$E_r \to {E_r}^{(h)}=E_r-\sigma_r h_r$ of the energy in this case. 
This yields:
\begin{equation}
\label{def_eqWh}
W_r^{(h)}=\frac{1}{2}\left(1-\sigma_r \tanh{[\beta_{|E_r|}(H_r+h_r)]}\right).
\end{equation}
The problem we face now is that an extra parameter $\beta_0$
appears, which governs the fate,
 under the influence of the applied field $h_r$, 
of configurations where $H_r=0$.
One  might think that the arbitrariness in the choice of $\beta_0$
 for a generic $(x,y)$-model could be resolved  by demanding that 
(\ref{def_Tdyn})  would give the same effective
 temperature for  all observables. This is not the case: in fact,
general properties of the spin-flip dynamics 
 we consider automatically guarantee that, for {\it any} given
value of $\beta_0$, our definition (\ref{def_Tdyn}) of
$\beta_{\rm dyn}=1/T_{\rm dyn}$
does not depend on the  observable  chosen to evaluate
the corresponding correlation and response functions
since it can be re-expressed as
\begin{equation}
\label{res_TdynvsW}
\beta_{\rm dyn}=\frac{\langle \sigma_r
(\partial_h  W_r^{(h)})_{h=0}\rangle}{\langle 
W_r \rangle}
=\frac{\langle 2 \beta_{|E_r|}W_r(1-W_r) \rangle}{\langle
 W_r \rangle},
\end{equation}
where the average is over all lattice sites   \cite{NOTE0}.

The way out of this quandary is deceptively simple: as for
the Glauber model, where $\beta_0$ is of course 
equal to the inverse of the true temperature,
one has to tune $\beta_0$ such that any
FDT measurement with a probing field yields precisely 
$\beta_0$ as the effective temperature. That is to say, for any
$(x,y)$-model, $\beta_0$ is the  solution of
\begin{equation}
\label{magic}
\beta_{\rm dyn}(\beta_0)=\beta_0.
\end{equation}
Because $\beta_0$ only appears, in Eqs.(\ref{def_Tdyn})-(\ref{res_TdynvsW}),
in the initial condition
$R(t,t^+)$ of the response function \cite{NOTE1},
the dependence of $\beta_{\rm dyn}$ on $\beta_0$ is simply {\it linear},
and the solution of (\ref{magic}) is therefore {\it unique}.

The reason why Eq.(\ref{magic}) is the correct prescription is a profound,
geometric one, and is incidentally the same as for the Glauber model:
this choice of $\beta_0$ actually maximizes the entropy of 
the spin configurations.
To understand this point, recall that at equilibrium
 the canonical Botzmann-Gibbs distribution can also be obtained from
the microcanonical ensemble under an information-theoretic,
 maximum-entropy condition \cite{JAYNES57},
where the Lagrange multiplier $\beta_{\rm geo}$ 
(identified with the inverse temperature in units where $k_B=1$)
 imposing the mean value of the energy 
  reweights with an exponential prefactor
$\propto e^{\beta_{\rm geo}E}$ the bare probabilities 
of the spin configurations.
Then, for any $(x,y)$-model on a finite system
a steady-state is always attained, and it is perfectly legitimate 
to define a microcanonical entropy $S_{\rm mic}(M,E) \equiv \ln{\Omega(M,E)}$
where the density of states $\Omega(M,E)$ simply 
counts the number of configurations having a prescribed
magnetization $M$ 
and energy $E$
as reached by the spin-flip dynamics.
But in this ``microcanonical'' ensemble,
 the only spin-flip  allowed are those which do not change the energy,
that is the ones corresponding to $W(0)=\frac{1}{2}$, a value common
to {\it all} $(x,y)$-models. In other words, the influence of 
distinct $(x,y)$-values can only be felt if one lets the energy fluctuates,
in which case --- and for exactly the same  reasons as
above --- one should expect to actually  observe the configurations with
an effective reweighting $\propto e^{\beta_{\rm geo} E}$,
leading to a ``canonical'' entropy $S_{\rm can}$.
Now, if one works with a non-zero external field $h$
in the microcanonical ensemble
(in particular around the most probable values
of $M$ and $E$),
  to lowest-order in $h$, once again,
only the  spin-flips with $H_r=0$ will be allowed,
with a probability (\ref{def_eqWh}) which can be rewritten as
$W^{(h)}(0)=\frac{e^{- \sigma_r \beta_0 h}}{e^{  \beta_0 h}+
e^{- \beta_0 h}}$.
This expression shows that a consistent reweighting
can be achieved if and only if $\beta_{\rm geo}=\beta_0$, an
equality that (\ref{magic}) also recovers if one restricts the average to
zero local-field configurations.

\begin{figure}
\includegraphics[width=8cm]{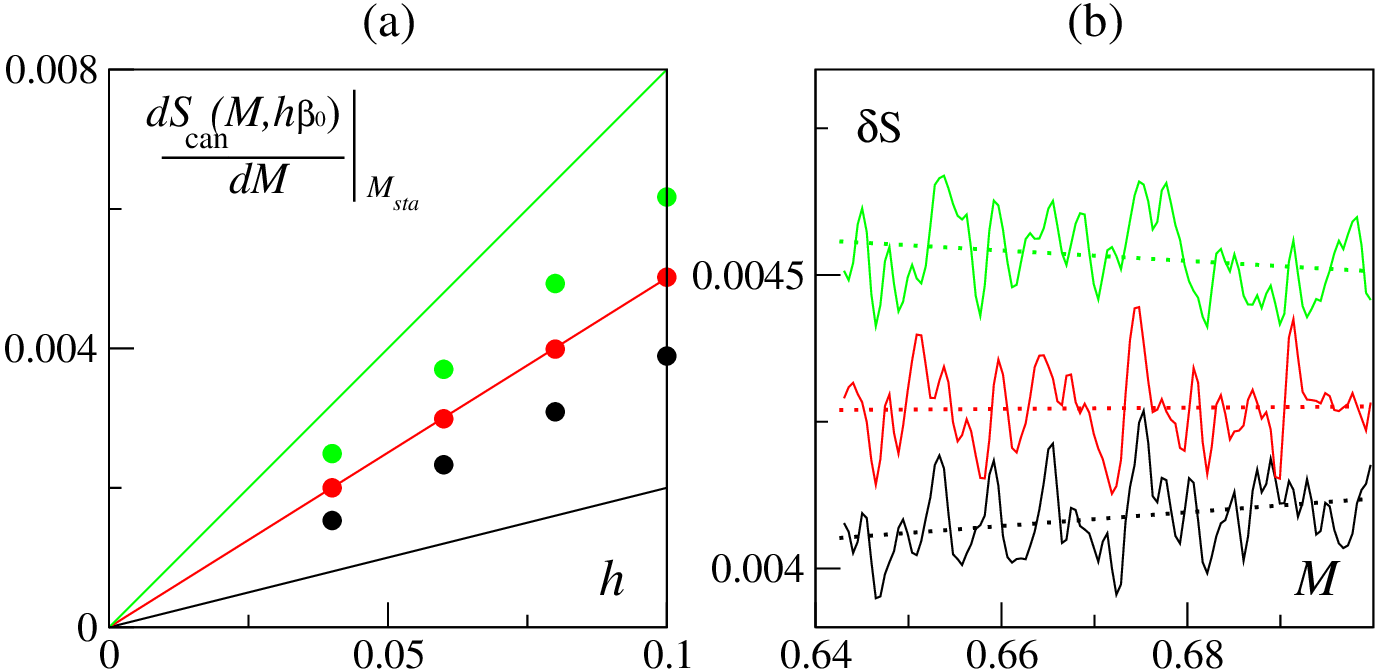}
\caption{(a): for the linear model at $x=0.1$ (disordered phase),
$\partial S_{\rm can}(M,h\beta_0)/\partial M$ evaluated at
$M_{\rm sta}=0$ for various $h$ values for $\beta_0=0.02$, 0.0501,
and 0.08 (from bottom up). It is only equal to $h\beta_0$ (solid lines)
for the middle value,  equal to $\beta_{\rm dyn}$.
(b) $\delta S$ vs $M$ for the critical majority
model ($x_{\rm c}\simeq 0.850$) 
with $\beta_0=0.365$, 0.415, and 0.465 from bottom up
($h=0.1$, $M_{\rm sta}\simeq 0.666$, dotted lines are linear fits).
Stationarity is achieved for the middle value, equal to $\beta_{\rm dyn}$.}
\label{f2}
\end{figure}

To check the validity of the above scenario, we have computed
the density of states (reweighted) using an efficient algorithm \cite{HUELLER}
which allows to determine, with high accuracy, the ratios 
$\frac{\Omega(M',E')}{\Omega(M,E)}e^{\beta_0(E'-E)}$ 
for neighboring (i.e. connected by a single
spin-flip) magnetization and energy levels.
This method is applicable anywhere in the $(x,y)$-plane, except 
of course  for $y=1, 1/2 \le x \le 1$ when no fluctuations
are present. It produces, in particular, 
fast and accurate estimates
by restricting the calculations to a narrow
interval centered around the stationary (and most probable) values
$M_{\rm sta}$ and $E_{\rm sta}$, with or without an external field. 
The above ratios also provide a direct access to an estimate
of $\frac{\partial S_{\rm can}(M,\beta_0 h)}{\partial M}$
which --- if the above effective
thermodynamic picture is valid --- 
should be equal to $h \beta_{\rm geo}$ in the disordered phase 
(where for any system size $M_{\rm sta}=0$). 
This is indeed verified for sufficiently small $h$ and, 
most importantly, it is only so
when one has tuned $\beta_0$ according to (\ref{magic}) 
(Fig.~\ref{f2}a).

In the ordered phase or along the critical line, finite-size
fluctuations of the energy
bring in another term (stemming from the differentiation of
 $e^{\beta_{\rm geo}E^{(h)}}$ in the effective occupation probabilities)
 which cannot be evaluated simply with the present method.
But one can nevertheless have an indirect access 
to the dependence of the absolute entropy
on $\beta_0$ by working, with and without magnetic field, 
along the magnetization direction in an 
interval comprising both $M^{(0)}_{\rm sta}$ and  $M^{(h)}_{\rm sta}$,
because this last quantity depends on $\beta_0$.
In particular, if the selected value of $\beta_0$ is the extremalizing one,
the difference 
$\delta S(M)=\frac{\partial S_{\rm can}(M,E,\beta_0 h)}{\partial M}-
\frac{\partial S_{\rm can}(M,E,0)}{\partial M}$ should be as flat as possible.
This is confirmed in Fig.~\ref{f2}b for the critical point of
the majority model.

\begin{figure}
\includegraphics[width=6.7cm]{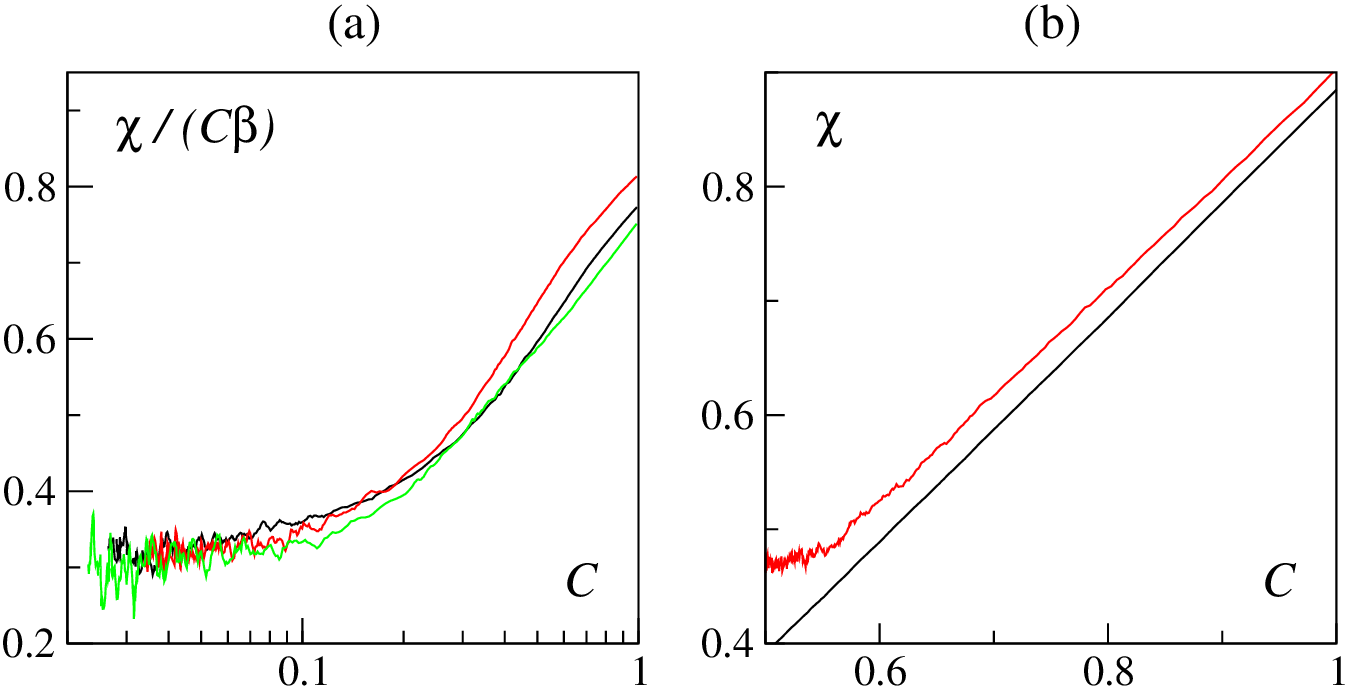}
\caption{(a) Parametric plot of
 $\chi/(C \beta)$ vs. $C$ for the Glauber, majority, and extreme models
at their critical point, where  $C=C(t,t')$ is the spin autocorrelation
function, and $\chi(t,t')=\int_0^t du R(u,t')$ is the integrated response
following the application, on random initial conditions, 
of a bimodal random field of magnitude $h \approx 0.05$ 
switched off at time $t=50$ (single runs on lattices of $8192^2$ spins
with periodic boundaries, data shown for $t'\in [t,3000]$
For small $C$, all curves level off around 
$X_{\infty}[M]=0.33(2)$.
(b) Short-time part of the $\chi-C$ plot
for the local magnetization (bottom) and the local energy
(top) in the voter model. Both curves have slope 
$\beta_{\rm voter}=\frac{\ln{3}}{4}$.}
\label{f3}
\end{figure}

To sum up, our nominal temperature is unique and can be determined in two
equivalent ways for any $(x,y)$ model (see Fig.~\ref{f1}b for an example).
We note in passing that two models sharing the same stationary energy
(e.g. Glauber and linear for which this quantity is known exactly
\cite{TBP}) are usually not at the same effective temperature.
Of particular interest is the non-monotonous variation
of the temperature $T_{\rm c}$ along the critical line  (Fig.~\ref{f1}c).
Once  $T_{\rm c}$ determined, one can
study the long-time behavior of $X$. 
Using a now-standard method \cite{BARRAT98},
the parametric representation of Fig.\ref{f3}a yields the universal limit
$X_{\infty}=0.33(2)$. This value while in disagreement with \cite{GODRECHE_2},
has also been obtained independently in \cite{BERTHIER_X,CHATELAIN},
and is consistent with the two-loop RG results of \cite{X_EPS}.

Further (analytical) insight can be gained for the
linear and critical voter model:
along the line $y=2 x$,
 all correlation functions do not couple to higher-order ones,
and verify diffusion  equations.
Using  standard Laplace-Fourier methods much as for the Glauber 
 $1d$-Ising chain, one can not only calculate exactly the stationary 
energy, but also obtain many exact results for the correlation 
and response functions. Skipping all technical details \cite{TBP},
the two-time two-point 
correlation function $C_r(t,t')=\langle \sigma_0(t) \sigma_r(t') \rangle$,
is compactly expressed in terms of a double Laplace transform
$\widehat{C}_r(s,s')=\frac{1}{s \hat{q}_0(s/2)}
\frac{\hat{q}_r(s/2)-\hat{q}_r(s')}{s'-s/2}$
($s \leftrightarrow t, s' \leftrightarrow t'$), where 
$q_r(t)=e^{-(1 -2 x)t}p_r(2 x t)$, $p_r(t)$ being the probability that a
 simple random walk goes from $0$ to $x$ during a length of time $t$ 
($p_0(t)=e^{-t}I^2_0(t/2) \simeq 1/(\pi t)$, 
where $I_0$ is a modified Bessel function).
Specializing to the autocorrelation function $C(t,t') \equiv C_0(t,t')$,
this gives 
for $x<1/2$ when $t$ is large and $t'-t$ arbitrary
$C(t,t') \simeq 1-\int_0^{t'-t}\!du  q_0(u)/\hat{q}_0(0)$, 
a function of 
$t'-t$  decaying exponentially on a timescale $\tau_x=1/(1-2 x)$ .
At the voter critical point, both $\hat{q}_0(0)$ and $\tau_x$ diverge,
and the previous expression crosses over to
 a non-trivial two-time scaling form, with 
$C(t,t') \simeq 1-{(\ln{t})}^{-1}\int_0^{t'-t}du p_0(u)$ 
in the first temporal
regime $t \gg 1, t'-t \sim {\cal{O}}(1)$ , which matches at large
time differences the  result $C(t,t')={(\ln{t})}^{-1}\ln{(\frac{t'+t}{t'-t})}$
valid in the second regime $t,t'-t \gg 1$.
As for the autoresponse function $R(t,t')$
 it simply reads $R(t,t')=q_0(t'-t)R(t,t^+)$.
 For $x<1/2$, $R(t,t^+)$ tends to a finite and non-zero
constant, and an effective FDT is obeyed at all times, with $X(t,t')=1$.
For the voter model,
$R(t,t^+) \simeq (\pi/\ln{t})[\beta_0-\frac{3 c_4}{8}(\beta_0-\beta_1)]$
in the long-time limit, where the numerical
 constant $c_4$ is such that $\langle
\prod_{\mu=1}^4 \sigma_{\mu}(t)\rangle \simeq 1-c_4 \pi/\ln{t}$.
Hence (\ref{res_TdynvsW})
 reads $\beta_{\rm dyn}(\beta_0)=\beta_0-\frac{3}{8}(\beta_0-\beta_1)c_4$, and the prescription (\ref{magic}) is realized when
 $\beta_0=\beta_1$, thus
bypassing the evaluation of $c_4$. Therefore the strength of 
interfacial noise driving the voter order-disorder
 transition can be unambiguously associated with 
a  temperature exactly given by $1/T_c=\frac{\ln{3}}{4}=0.27465\dots$, 
and the FDT ratio 
$X(t,t')\simeq\frac{p_0(t'-t)}{p_0(t'+t)+p_0(t'-t)} \to X_{\infty} =\frac{1}{2}$
\cite{NOTE3}.
While $T_{\rm c}$ is easily measured for both $M$
and $E$ and found equal to the exact value above  (Fig.~\ref{f3}b), 
the long-time regime is difficult to reach numerically.
For the magnetization, 
we find $X_{\infty}[M] =0.50(5)$ (not shown), which gives an indication of
the accuracy of the Ising value above.
For the energy, however, the signal is too weak to estimate  $X_{\infty}[E]$
so that we cannot test the conjecture \cite{BERTHIER_X} that 
$X_{\infty}$ is the same for all observables, which seems unlikely to us
given that critical static amplitudes ratios usually depend 
on the observable considered. 

Summarizing, we have shown how to measure the ``strength of noise''
via a {\it bona fide} thermodynamic temperature in kinetic spin models
and confirmed the universality of the FDT ratio  $X_{\infty}$ at critical
points. Ongoing work aims at extending our method to even more general
nonequilibrium systems such as chaotic map lattices.

F.\ S.\ thanks Conacyt~(M\'exico) for financial support.


\begin{thebibliography}{99}
\bibitem{NOTEXX} The  limit $t \to \infty$ is somewhat
formal: after some microscopic time (\ref{def_Tdyn}) becomes actually 
time-independent.
\bibitem{NOTE0}
The proof is easy, if formal. For instance, if
 $C_M(t,t')=\langle \sigma_r(t) \sigma_r(t') \rangle$ is the
 autocorrelation   of the (local) magnetization, and 
$C_{O_n}(t,t')=\langle O_n(t)  O_n(t') \rangle$
that of an observable $O_n=\prod_{k=1}^n\sigma_{r_k}$,
then 
$\partial_{t'} C_{O_n}(t,t')=-2\sum_{k}\langle O_n(t) O_n(t') W_{r_k}(t')
\rangle$ tends to  
$-2\sum_{k}\langle W_{r_k}(t)\rangle= n \partial_{t'} C_M(t,t')|_{t'=t^+}$
as $t'\to  t^+$ (due to  the continuity of the master equation,
 ${\sigma_{r_k}}^2(t)=1$ and 
translation invariance). Similarly, the corresponding
response functions are related by $R_{O_n}(t,t^+)=n R_M(t,t^+)$,  and
the factor $n$ thus cancels  when calculating  
the ratio appearing in (\protect{\ref{def_Tdyn}}).
\bibitem{NOTE3} These equalities are valid in any space dimension,
so that $X_\infty=\frac{1}{2}$ for the voter class as defined
in \cite{VOTPRL}.
\end{thebibliography}
\end{document}